\documentclass[aps,prl,reprint,superscriptaddress]{revtex4-2}

\usepackage[version=3]{mhchem} 
\usepackage{graphicx}          
\usepackage{amsmath,amssymb}   
\usepackage{xcolor}            
\usepackage{siunitx}           
\usepackage{booktabs}          
\usepackage{multirow}          
\usepackage{array}             
\usepackage[hidelinks]{hyperref} 

\begin{document}

\title{Decoupling Bias Position and Timing in Metadynamics}

\author{Konstantin Stracke}
\affiliation{School of Physics, Chemistry and Earth Sciences, Adelaide University, City East Campus, Adelaide, SA 5005, Australia}
\author{David M. Huang}
\affiliation{Laboratoire de Physique, \'Ecole Normale Sup\'erieure, Universit\'e PSL, CNRS UMR 8023, 75005 Paris, France}
\author{Jack D. Evans}
\email{j.evans@adelaide.edu.au}
\affiliation{School of Physics, Chemistry and Earth Sciences, Adelaide University, City East Campus, Adelaide, SA 5005, Australia}

\begin{abstract}
Metadynamics deposits biases at the instantaneous collective variable (CV) position at fixed intervals, so the bias tracks the dynamics rather than the underlying landscape. We decouple bias position and timing using two approaches: Mean-CV deposits at the interval-averaged position and Arc-Length per unit path length in CV space. For a water bend that oscillates faster than it thermalizes, conventional metadynamics overweights turning points, whereas Arc-Length removes this residence-time weighting and Mean-CV fills the well center early. More broadly, decoupling where and when hills are added opens a new avenue for discrete biasing.
\end{abstract}

\maketitle

Molecular dynamics is largely governed by the potential energy surface and the dynamics with which it is explored, but the two need not coincide.
Simulations uphold ergodicity and equipartition only in the limit of long time scales, not instantaneously and not locally.
Metadynamics simulations were developed to overcome free energy barriers and accelerate sampling on the free energy landscape\cite{laio2002escaping}.
Since its introduction, metadynamics has deposited bias at the instantaneous collective variable (CV) position, typically after a fixed number of simulation steps, irrespective of the local dynamics.
For Gaussian kernels, the metadynamics bias potential is
\begin{equation}
V(s,t)=\sum_{i=1}^{N(t)}
w_i
\exp\left[
-\frac{\left(s-s_i\right)^2}{2\sigma^2}
\right],
\label{eq:metadynamics}
\end{equation}
where $w_i$ is the height and $s_i$ the center of the $i$-th Gaussian, $\sigma$ the Gaussian width, $s$ the CV coordinate, $t$ the simulation time, and $N(t)$ the number of Gaussians deposited up to time $t$.
Over the past two decades, significant methodological advancements have optimized almost every parameter of this deposition equation except $s$, the position.
Well-tempered metadynamics dynamically scales the bias height ($w_i$) to ensure convergence of the free energy\cite{barducci2008well}, while dynamically adapted schemes alter the Gaussian width ($\sigma$) based on the local diffusivity\cite{branduardi2012metadynamics}.
Notably, frequency-adaptive metadynamics introduces temporal adaptation by dynamically adjusting the deposition stride based on the accumulated bias, rather than the underlying potential energy surface\cite{wang2018frequency}.
Related approaches have reformulated bias construction through on-the-fly probability estimation (OPES), mobile walker-centered Gaussians (Flying Gaussian) or adaptively learned local CV representations (reconnaissance metadynamics)\cite{invernizzi2020rethinking,sucur2016sampling,tribello2010self}.
However, for successive discrete kernels deposited within a fixed CV space, the kernel center has remained tied at deposition to the instantaneous trajectory position.

In this work, we introduce a new approach to metadynamics through positional and temporal decoupling.
We achieve positional decoupling by depositing the bias at the average collective variable position sampled since the previous deposition (Mean-CV).
On the other hand, temporal decoupling replaces the conventional time-based deposition schedule with a distance-based criterion in collective variable space (Arc-Length).
We demonstrate these approaches on the H--O--H bend angle of water, which oscillates faster than it can exchange energy with its surroundings, so that the dynamics and underlying thermodynamics do not coincide.
The choice of deposition rule matters most when only a limited number of biases can be afforded, as in ab initio metadynamics\cite{iannuzzi2003efficient,bonati2018silicon}, because the first biases then shape the free energy estimate.

Conventional metadynamics deposits bias at the instantaneous CV position at fixed time intervals:
\begin{equation}
s_i=s(t_i);
\qquad
t_i=i\tau,
\label{eq:classical_deposition}
\end{equation}
where $t_i$ and $s_i$ denote the deposition time and position of the $i$-th bias, respectively, and $\tau$ is the deposition stride.
Mean-CV retains this schedule but replaces the instantaneous position by the mean over the preceding time interval,
\begin{equation}
s_i=\bar{s}(t_i);
\qquad
t_i=i\tau,
\label{eq:Mean-CV_deposition}
\end{equation}
where $\bar{s}(t_i)$ denotes the average over the CV positions accumulated during the preceding deposition interval:
\begin{equation}
\bar{s}(t_i)=\frac{1}{\tau}\int_{t_{i-1}}^{t_i}s(t)\,\mathrm{d}t.
\label{eq:Mean-CV_deposition_2}
\end{equation}
Arc-Length, on the other hand, retains the instantaneous deposition position,
\begin{equation}
s_i=s(t_i),
\label{eq:Arc-length_s}
\end{equation}
but replaces the fixed temporal stride with a fixed distance $L_0$ traveled in CV space.
This results in variable deposition times,
\begin{equation}
t_i=\min\left\{t:L(t)-L(t_{i-1})\geq L_{0}\right\},
\label{eq:Arc-length_deposition}
\end{equation}
where the traveled distance is accumulated from the preceding deposition at $t_{i-1}$ until it reaches $L_0$ at $t_i$.
The cumulative distance traveled in CV space is defined as
\begin{equation}
L(t)=\int_{0}^{t}\left|\dot{s}(t')\right|\mathrm{d}t',
\label{eq:t_arc}
\end{equation}
where $\dot{s}(t')=\mathrm{d}s/\mathrm{d}t'$ describes how the CV changes with time, and its absolute value ensures that motion in either direction contributes positively to $L(t)$.
At long times, all three deposition schemes give similar distributions of bias centers, as shown in the Supplemental Material\cite{SupplementalMaterial}\nocite{berendsen1981interaction,wilson1980molecular,fixman1974classical}, Fig.~S1.

\begin{figure}
\begin{center}
\includegraphics[width=\linewidth]{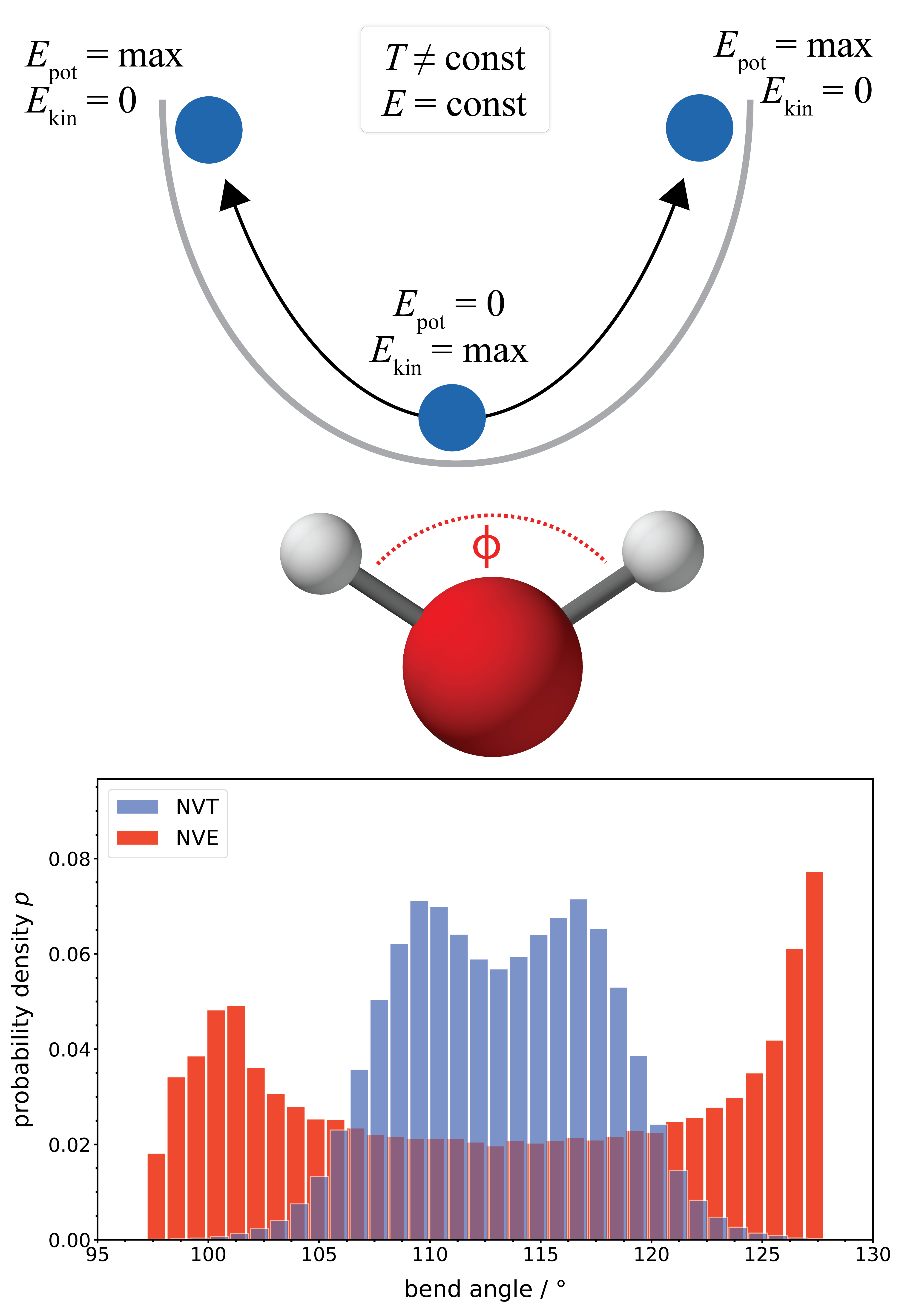}
\caption{Top: schematic of an oscillator at constant total energy, in which kinetic ($E_{\text{kin}}$) and potential ($E_{\text{pot}}$) energy interconvert over each period, so that the microcanonical ($E=\text{const}$, $T\neq\text{const}$) rather than the canonical ($T=\text{const}$, $E\neq\text{const}$) ensemble is sampled. Middle: the H--O--H bend angle of water used as the CV. Bottom: bend-angle distributions from unbiased molecular dynamics in the NVE ensemble and in the NVT ensemble with a weakly coupled thermostat (both at \SI{200}{K}, with a \SI{0.25}{fs} timestep).}
\label{result:schematic}
\end{center}
\end{figure}

To challenge the different metadynamics schemes, we consider a harmonic oscillator at fixed total energy (i.e.\ microcanonical conditions) following deterministic periodic motion, for which the position distribution follows the classical arcsine law, diverging at the turning points:
\begin{equation}
p(x)\propto \frac{1}{|v(x)|} \propto \frac{1}{\sqrt{E-U(x)}},
\end{equation}
where $v(x)$ is the instantaneous velocity and $U(x)$ the potential energy.
This residence time weighting is particularly consequential for metadynamics, in which bias is deposited according to the visited distribution of a CV.
If the thermostat coupling is weak, relative to the intrinsic oscillation period, the dynamics follow quasi-microcanonical motion over the timescale relevant for bias deposition (see Fig.~\ref{result:schematic}).

In this regime, biases are deposited according to the instantaneous residence time distribution, which remains arcsine-like and weighted toward the turning points, rather than according to the canonical probability density.
This is not an artifact of imperfect thermostatting.
In real systems, many modes are not fully thermalized when heat exchange with the environment is negligible or slower than the dynamics of interest.
In these cases fast local modes can outpace global thermal relaxation.
A separation of timescales can arise in gas-phase molecules\cite{stuart1996molecular} and, for stiff intramolecular modes, also in condensed matter\cite{hu2012temperature}.
The H--O--H bend of water is a case in point, with a period of \SI{20}{fs} (\SI{1640}{\per\centi\metre}), several times shorter than the $170\pm30$~fs lifetime of the first excited bend level in the liquid\cite{huse2005ultrafast}.

All water metadynamics and NVT simulations were performed with the YAFF (Yet Another Force Field) package\cite{verstraelen2022yaff} at \SI{200}{K} with a timestep of \SI{0.25}{fs}.
The temperature was controlled by a Nosé--Hoover chain thermostat of length three with a relaxation time of \SI{0.1}{ps}\cite{nose1984molecular, nose1984unified, hoover1985canonical, martyna1992nose}.
For the Conventional, Mean-CV, and Arc-Length simulations, we employed the `bias-deletion' well-tempered metadynamics scheme\cite{stracke2026bias}, which deletes all accumulated biases after a given number of depositions.
This approach allows for the resampling of the free energy surface multiple times within one simulation.
For example, with a 10-bias deletion window, the average deposition histogram of the first 10 biases can be reconstructed from many repetitions.
Alternatively, this approach may be viewed as an ensemble of many short simulations with the associated advantages and disadvantages\cite{stracke2026bias}.
Deletion windows of 10, 25, 50, 75, 100, 150, 250, 1000, 5000, \num{10000}, and \num{20000} biases were used, and the bias factor was set to 10.
Irrespective of the deletion window, the first \num{100000} deposited biases of each simulation were analyzed.
The water model was described by the SPC/Fw\cite{wu2006flexible} force field.
For all schemes, the initial bias height and the Gaussian width were set to \SI{0.1}{\kilo\joule\per\mole} and \SI{1}{\degree}, respectively.

As an illustrative comparison, the model for water was simulated using molecular dynamics and the different metadynamics schemes.
The molecular dynamics simulations were performed in both the NVE and NVT ensembles.
For the NVT simulation, the thermostat relaxation time was set to approximately five times the bend period (Fig.~\ref{result:schematic}).

\begin{figure}
\begin{center}
\includegraphics[width=\linewidth]{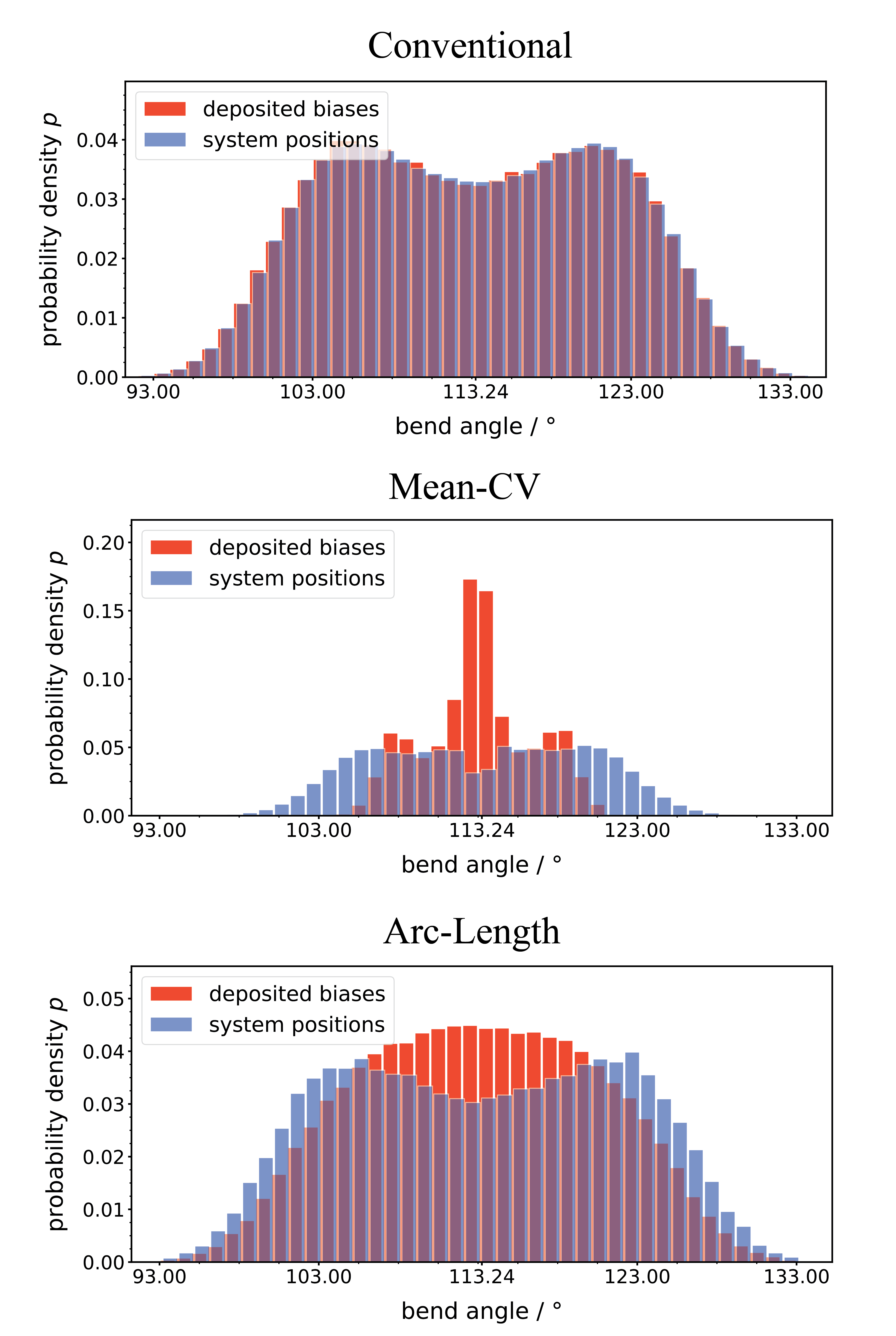}
\caption{Bend-angle distributions sampled by the system (blue) and distributions of deposited bias centers (red) for Conventional, Mean-CV, and Arc-Length metadynamics.}
\label{result:MTD}
\end{center}
\end{figure}

Within the well-tempered framework, we compare conventional metadynamics (Conventional) with Mean-CV and Arc-Length by the mechanism through which each fills the free energy surface.
As expected, Conventional metadynamics reproduces the underlying molecular dynamics probability distribution, depositing bias preferentially near the turning points (Fig.~\ref{result:MTD}).
Contrastingly, Mean-CV fills the center of the well from the outset by depositing the bias at the average CV position.
As the central region becomes increasingly biased, the trajectory becomes temporarily confined to one side of the landscape, where the local average CV shifts and subsequent biases continue to accumulate there.
Once sufficient bias has accumulated, the system transitions to the opposite side, and the process repeats.
This alternating filling mechanism produces the characteristic triple-peaked bias distribution and corresponding double-arcsine trajectory distribution seen in Fig.~\ref{result:MTD}.

We find that the Arc-Length scheme recovers the corresponding NVT-like distributions, as shown in Fig.~\ref{result:MTD}.
Notably, the arcsine-like trajectory distribution is retained, but the bias smoothly fills the free energy well.
As the bias placement is coupled to CV space and not time, the deposited bias no longer reflects how long the system resides at each CV value (Fig.~\ref{result:MTD}).

\begin{figure}
\begin{center}
\includegraphics[width=\linewidth]{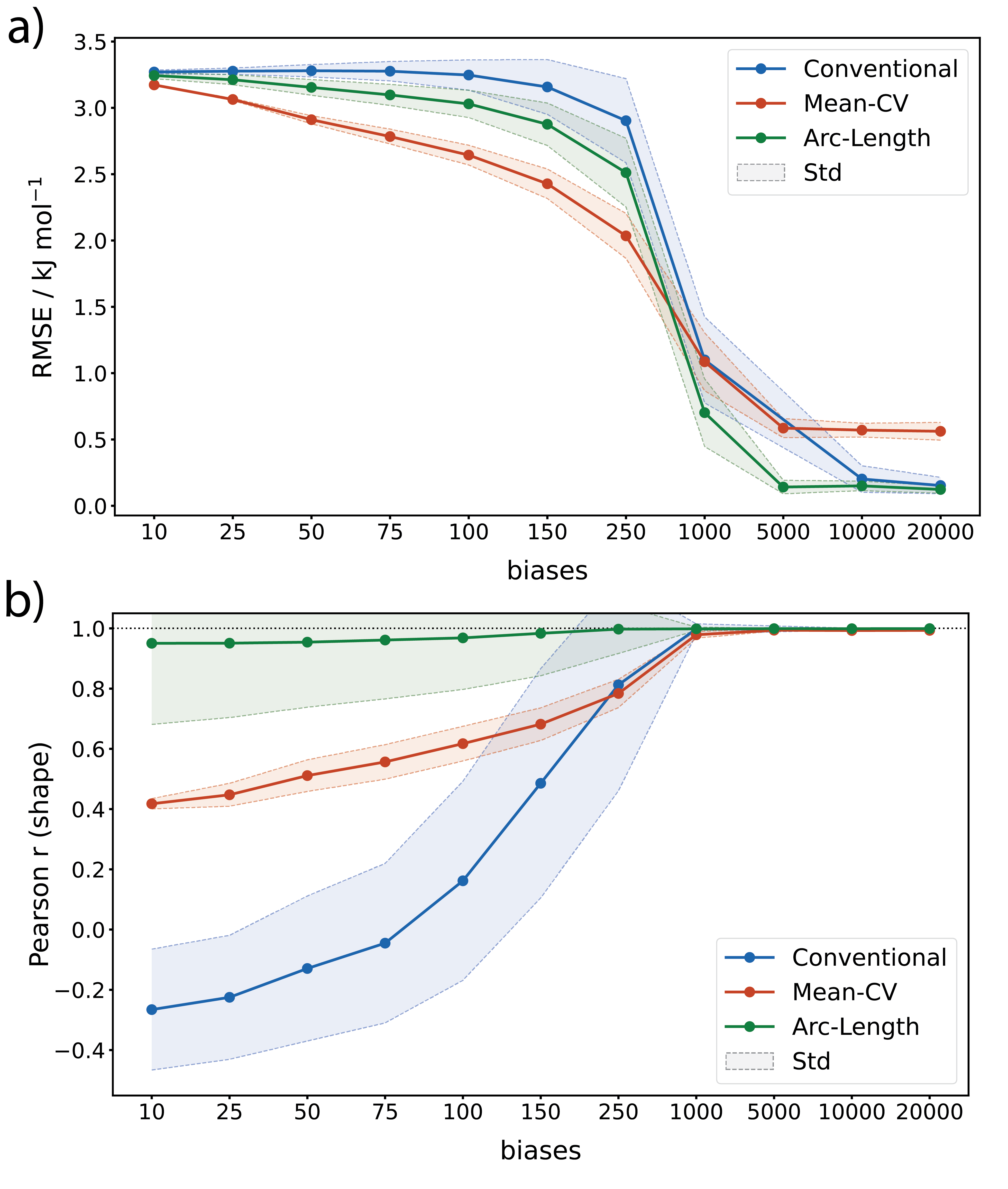}
\caption{(a) RMSE and (b) Pearson $r$ of the reconstructed free energy with respect to the analytical reference (Supplemental Material, Eq.~S2) within $\pm15^\circ$ of the equilibrium bend angle, for Conventional, Mean-CV, and Arc-Length metadynamics as a function of the number of biases per deletion window. Pearson $r$ quantifies shape similarity. Solid lines and shaded regions indicate the full-trajectory estimate and the standard deviation across deletion windows, respectively.}
\label{result:quantitative}
\end{center}
\end{figure}

In addition to this qualitative examination of the filling mechanisms, we also considered the statistical error and shape similarity when sampling.
Since Conventional and Mean-CV deposit at a fixed stride of 1000 MD steps (\SI{0.25}{ps}), whereas the Arc-Length deposition interval ($L_0=\SI{500}{\degree}$) varies with the speed of the CV, we compared the methods as a function of the number of deposited biases to ensure a fair comparison (see Figs.~S5 and S6 for the corresponding comparison in simulation time).
The root-mean-square error (RMSE) with respect to the analytical free energy (Supplemental Material, Eq.~S2) and the Pearson $r$ value were calculated for deletion windows containing different numbers of biases (Fig.~\ref{result:quantitative}).
The Pearson $r$ value quantifies shape similarity independent of the overall scale and additive offset of the reconstructed bias, analogous to its use as a map-model correlation coefficient in cryo-EM validation\cite{afonine2018new, kidmose2019namdinator}.
Values approaching unity therefore indicate that maxima, minima, and intermediate features occur at corresponding CV positions in the reference and tested profiles.
For $M$ evaluation points $s_k$ along the CV, let $x_k=F_{\mathrm{recon}}(s_k)$ and $y_k=F_{\mathrm{analytical}}(s_k)$, where $F(s)$ denotes the free energy profile as a function of the CV.
The Pearson correlation is
\begin{equation}
r=
\frac{\sum_k (x_k-\bar{x})(y_k-\bar{y})}
{\sqrt{\sum_k (x_k-\bar{x})^2}
 \sqrt{\sum_k (y_k-\bar{y})^2}}.
\label{eq:pearson_r}
\end{equation}
Figure~\ref{result:quantitative} immediately highlights the rapid convergence of Mean-CV.
By depositing the bias at the center of the potential well from the outset, the RMSE decreases substantially during the early stages of the simulation, yet the long-time convergence suffers.
Compared to the Conventional method, using the same deposition stride, Mean-CV converges faster; for 250-bias deletion windows, its RMSE is about \SI{30}{\percent} lower at the same simulation time.
However, convergence and correlation do not necessarily improve in parallel. For some bend angle ranges, faster convergence comes at the cost of lower correlation (Fig.~S2), whereas in most cases both faster convergence and higher correlation can be achieved simultaneously (Fig.~S4).

In contrast to Mean-CV, Arc-Length exhibits a high Pearson correlation over all studied ranges ($r\geq0.78$; Fig.~S2). Even early in the simulation, high shape similarity shows that the bias follows the contour of the potential well.
Arc-Length outperforms Conventional metadynamics at intermediate window sizes in both RMSE per deposited bias and functional similarity, and is never worse. Because the underlying free energy surface is sampled from the outset, no subsequent `self-healing'\cite{dickson2015mu} is required.

These conceptual filling mechanisms are independent of the well-tempered implementation and persist in non-well-tempered metadynamics (Fig.~S8), indicating their potential applicability to other adaptive biasing schemes.
Our results demonstrate that positional and temporal decoupling can accelerate convergence and promote uniform sampling of the underlying free energy surface.

\begin{figure}
\begin{center}
\includegraphics[width=\linewidth]{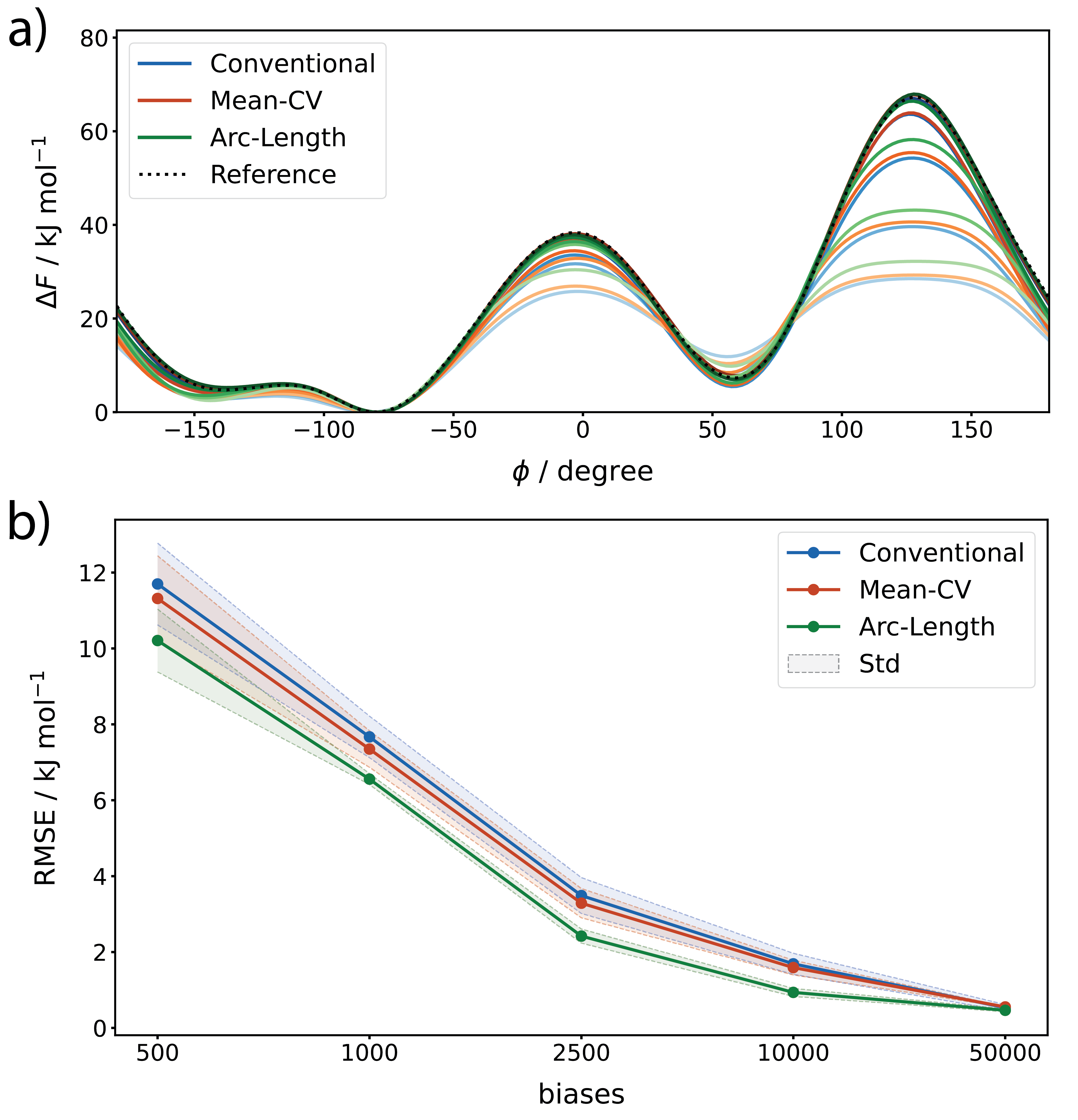}
\caption{Conventional, Mean-CV, and Arc-Length metadynamics results for alanine dipeptide: (a) one-dimensional Helmholtz free energy profiles projected onto $\phi$ and (b) RMSE as a function of the number of deposited biases. The profiles in (a) correspond to the different deletion windows, with lighter shades indicating fewer biases.}
\label{result:quantitative_alanine}
\end{center}
\end{figure}

Moving beyond the deliberately extreme water bend model, we put these same deposition strategies to the test on the prototypical alanine dipeptide.
The same simulation protocol was applied to alanine dipeptide at \SI{300}{K} using a \SI{2}{fs} timestep, a bias height of \SI{1.3}{\kilo\joule\per\mole}, a width of \SI{20}{\degree}, and the AMBER99SB-ILDN\cite{lindorff2010improved} force field.
Biases were deposited every 50 MD steps for Conventional and Mean-CV metadynamics, and every \SI{500}{\degree} of cumulative displacement of both CVs for Arc-Length metadynamics; Mean-CV used the circular mean of each dihedral, and Arc-Length the Euclidean path length in the $(\phi,\psi)$ plane.
Deletion windows of 500, 1000, 2500, \num{10000}, and \num{50000} biases were each analyzed over the first \num{100000} bias depositions and referenced against a Conventional metadynamics simulation with a deletion window of \num{100000} biases.
For the two-dimensional free energy surface the intermediate estimates do not appear shape-preserving, meaning that the relative locations and prominence of the principal minima, barriers, and connecting features change as the bias accumulates.
For example, the $500$-bias deletion window is not a scaled-down representation of the converged surface.
Therefore, instead of the Pearson $r$, the one-dimensional projected free energy profiles are shown directly in Fig.~\ref{result:quantitative_alanine}(a), with the corresponding two-dimensional bias-center distributions shown in Fig.~S9.
The RMSE analysis [Fig.~\ref{result:quantitative_alanine}(b)] shows that Arc-Length converges with fewer bias depositions, and Mean-CV only marginally, within the spread across deletion windows.

The deposition concepts therefore extend to more complex, anharmonic free energy landscapes defined by coupled CVs under canonical-ensemble sampling.
We show that deposition positions and timing intervals need not be coupled to instantaneous clock steps, so long as the resulting hill distribution is consistent with the properly reweighted underlying Boltzmann measure.

In summary, we introduced two general metadynamics strategies that decouple bias deposition from either the instantaneous system position or the physical simulation time.
These strategies are independent of, and can be combined with, bias modification schemes such as well-tempered metadynamics.
They reveal that the deposition strategy fundamentally determines the bias-filling mechanism, and that controlled decoupling enables faster convergence and sampling that faithfully reflects the underlying energy landscape.

We envisage future developments that combine positional and temporal decoupling and tailor adaptive biasing strategies to exploit their complementary advantages.
Such combinations could incorporate position- and time-adaptive biases driven by Bayesian, machine-learned, or other statistical estimators.
Decoupling where and when bias is deposited provides a new design principle for developing faster and more robust enhanced sampling algorithms.

\textit{Data availability}---The simulation inputs, the metadynamics implementation, the deposited-bias logs, and the analysis scripts are openly available on Zenodo at \href{https://doi.org/10.5281/zenodo.24549519}{doi:10.5281/zenodo.24549519}.

\begin{acknowledgments}
J.D.E. is the recipient of an Australian Research Council Discovery Early Career Award (project number DE220100163) funded by the Australian Government. The Phoenix HPC service at Adelaide University is thanked for providing high-performance computing resources. This research was supported by the Australian Government's National Collaborative Research Infrastructure Strategy (NCRIS), with access to computational resources provided by Pawsey Supercomputing Research Centre through the National Computational Merit Allocation Scheme.
\end{acknowledgments}

\bibliography{main}

\end{document}